\documentclass[%
 reprint,
 amsmath,amssymb,
 aps,
]{revtex4-2}

\newcommand{\ms}[1]{\texttt{#1}}
\usepackage{float}

\newcommand{\bx}{\mathbf{x}}
\newcommand{\bq}{\mathbf{q}}

\usepackage{graphicx}% Include figure files
\usepackage{dcolumn}% Align table columns on decimal point
\usepackage{bm}% bold math
\usepackage{listings}
\usepackage{xcolor}

\definecolor{codegreen}{rgb}{0,0.6,0}
\definecolor{codeblue}{rgb}{0,0.0,0.6}
\definecolor{codegray}{rgb}{0.5,0.5,0.5}
\definecolor{codepurple}{rgb}{0.58,0,0.82}
\definecolor{backcolour}{rgb}{0.95,0.95,0.92}
\lstdefinestyle{mystyle}{
    backgroundcolor=\color{backcolour},   
    commentstyle=\color{codegreen},
    keywordstyle=\color{magenta},
    numberstyle=\tiny\color{codeblue},
    stringstyle=\color{codepurple},
    basicstyle=\ttfamily\footnotesize,
    breakatwhitespace=false,         
    breaklines=true,                 
    captionpos=b,                    
    keepspaces=true,                 
    numbers=left,                    
    numbersep=5pt,                  
    showspaces=false,                
    showstringspaces=false,
    showtabs=false,                  
    tabsize=2
}

\begin{document}

%\preprint{APS/123-QED}

\title{cuPSS: a package for pseudo-spectral integration of stochastic PDEs}% Force line breaks with \\

\author{Fernando Caballero}
 \affiliation{Department of Physics, Brandeis University, Waltham, Massachusetts 02453, USA}%Lines break automatically or can be forced with \\

 \email{fcaballero@brandeis.edu}
 \homepage{http://github.com/fcaballerop/cuPSS}
 \affiliation{Department of Physics, University of California Santa Barbara, Santa Barbara, CA 93106, USA}

\date{\today}% It is always \today, today,
             %  but any date may be explicitly specified

\begin{abstract}
A large part of modern research, especially in the broad field of complex systems, relies on the numerical integration of PDEs, with and without stochastic noise. This is usually done with eiher in-house made codes or external packages like MATLAB, Mathematica, Fenicsx, OpenFOAM, Dedalus, and others. These packages rarely offer a good combination of speed, generality, and the option to easily add stochasticity to the system, while in-house codes depend on certain expertise to obtain good performance, and are usually written for each specific use case, sacrificing modularity and reusability. This paper introduces a package written in CUDA C++, thus enabling by default gpu acceleration, that performs pseudo-spectral integration of generic stochastic PDEs in flat lattices in one, two and three dimensions. This manuscript  describes the basic functionality of cuPSS, with an example and benchmarking, showing that cuPSS offers a considerable improvement in speed over other popular finite-difference and spectral solvers.
\end{abstract}

\maketitle

%\tableofcontents

\section{Introduction}

Many interesting phenomena are described by stochastic partial differential equations when some fast degrees of freedom can be integrated out into noise on mesoscopic timescales. These include systems in fields as varied as fluid mechanics \cite{forster1977large}, ferromagnetism \cite{ma1975critical}, forms of phase separation \cite{hohenberg1977theory,tjhung2018cluster}, interface dynamics \cite{kardar1986dynamic,bray2001interface,caballero2018strong,besse2023interface}, soft and active matter \cite{wittkowski2014scalar,tiribocchi2015active,caballero2018bulk,caballero2022activity,chate2024dynamic,besse2022metastability}, systems biology \cite{lande2003stochastic,wilkinson2018stochastic}, finance \cite{mastromatteo2014anomalous,toth2011anomalous}, and many others. These equations, in what will interest us below, can be generally written as a function of a field $\phi(\bx,t)$, as
\begin{equation}\label{eq:base}
    d\phi = F[\phi]dt + \sqrt{2D}dW,
\end{equation}
where $F[\phi]$ is a function describing the deterministic evolution of $\phi$, while $dW$ is a Wiener process. I assume that all differential equations are written calculated in the Ito formulation when writing any discretization for them, which allows a more natural description of the noise in discrete time \cite{kloeden1992numerical}. I also allow the strength of the noise $D$ to be a differential operator, which will be useful for describing different types of noise \cite{hohenberg1977theory}.

Despite the rich theoretical machinery to analyze the properties of this type of equations, their stochastic nature, and usually complex form of their deterministic evolution, makes numerical methods a powerful tool in investigating them. These methods offer a quick insight into their behavior, as well as a way to test hypotheses about a model prior to investing more effort into different approaches.

\subsection{Pseudo-spectral integration}

Of the plethora of methods to numerically integrate an equation as Eq. \ref{eq:base}, spectral methods offer the advantage of higher numerical stability to space discretization, especially when the deterministic part of the evolution $F[\phi]$ has terms with many gradients. Writing the Fourier transform of $\phi(x,t)$ in space as $\tilde\phi(\bq,t) = (2\pi)^{-d/2}\int d\bq e^{-i\bq\cdot \bx}\phi(\bx,t)$, Eq. \ref{eq:base} as can be rewritten as
\begin{equation}\label{eq:base2}
    \partial_t\tilde\phi(\bq,t) = \tilde F [\tilde\phi] + \sqrt{2\tilde D}\tilde\eta,
\end{equation}
where I have dropped the differential form of the starting equation, and rewritten the noise as $\tilde\eta$, now defined with an average of $0$ and a variance
\begin{equation}
    \langle \eta(\bq,t)\eta(\bq',t'\rangle = (2\pi)^d\delta^{(d)}(\bq+\bq')\delta(t-t').
\end{equation}

The noise strength is written with a tilde because, as mentioned above, it is not necessarily a constant. For instance, a common form of Eq. \ref{eq:base2} is a continuity equation for a mass conserving field, $\partial_t\phi = -\nabla\cdot \vec J$, for some current $\vec J$. A noise term to this system is therefore applied to the current, so that $\vec J = \vec J_d + \vec\Lambda$, where $\vec J_d$ is the deterministic part of the current, and $\vec\Lambda$ is a vectorial noise. A simple calculation can show that, in Fourier space, this noise can be written as in Eq. \ref{eq:base2}, where $\tilde D = D q^2$ \cite{hohenberg1977theory}.

A drawback of spectral methods is their inefficiency when it comes to computing nonlinear terms in $F[\phi]$, since polynomial nonlinearities of order $n$ will involve convolutions that take $O(N^n)$ steps to compute, where $N$ is the number of steps in which space is discretizes. A solution for dealing with these types of nonlinearities in a more efficient way is to use the so-called pseudo-spectral methods \cite{fornberg1998practical}, where the nonlinearities are instead calculated in real space, while gradients are calculated in Fourier space. This offers a considerable speedup, despite the need to transform fields back and forth between real and Fourier spaces, since there are fast Fourier transform algorithms (commonly known as FFT algorithms \cite{FFTW05}) that work at a speed of $O(N\log N)$. 

For instance, let's consider a simple example of a deterministic evolution of a field $\phi(\bx,t)$ with cubic decay
\begin{equation}
    \partial_t\phi(\bx,t) = -\phi(\bx,t)^3.
\end{equation}
This equation can be readily written for the Fourier transform of $\phi(x,t)$, $\tilde\phi(\bq,t)$, with wavevector $\bq$,
\begin{equation}
    \partial_t\tilde\phi(\bq,t) = -\iint \frac{d\mathbf{k}d\mathbf{l}}{(2\pi)^{2d}} \tilde\phi(\mathbf{k},t)\tilde\phi(\mathbf{l},t)\tilde\phi(\bq-\mathbf{k}-\mathbf{l},t).
\end{equation}

Computing the right-hand side numerically in a system discretized in a lattice of size $N$ would thus require $O(N^3)$ calculations; $O(N^2)$ calculations for the convolution at each wavevector $\bq$. A pseudo-spectral method consists of viewing the previous equation in the following equivalent way
\begin{equation}\label{eq:cubicexample}
    \partial_t\tilde\phi(\bq,t) = \mathcal F\left[\mathcal F^{-1}\left[\tilde\phi(\bq,t)\right]^3\right],
\end{equation}
where $\mathcal F$ and $\mathcal F^{-1}$ are the Fourier transform and inverse Fourier transform operators respectively. This method will require only $O(N^2\log N)$ steps, since multiplication in real space is linear in the number of lattice points, and Fourier transforms can be computed at a faster speed than a naive convolution.

A pseudo-spectral method thus consists of numerically integrating Eq. \ref{eq:base2} in Fourier space, but calculating nonlinear terms by transforming each factor to real space, calculating the product in real space, and transforming the result back to Fourier space.

\subsection{De-aliasing}

One last detail must be mentioned when talking about these numerical methods, for which we need to introduce an explicit discretization of space. Consider a field $\phi(x,t)$, defined in a one-dimensional space of size $N$, discretized in $N/\Delta x$ points, each separated from the next at a distance of $\Delta x$. This induces a natural discretization of Fourier space in $N$ frequencies $q_i$ with a spacing of $\Delta q = 2\pi/(N\Delta x)$, where it is common to choose a frequency interval symmetric around $q=0$, such that $q_i\in(-\pi/\Delta x,\pi/\Delta x)$. Let us assume for simplicity, and without loss of generality, that $\Delta x = 1$. We then have a lattice of $N$ frequencies going from $-\pi$ to $\pi$. Let us now consider the discrete Fourier series of the field $\phi(x)$, which can be written $\phi(x) = \sum_{q_i} \tilde\phi(q_i) \exp(i q_i x)$. Consider a quadratic nonlinearity derived from this field
\begin{equation}
    \phi(x)^2 = \sum_{q_i,q_j}\tilde\phi(q_i)\tilde\phi(q_j)\exp(i(q_i+q_j)x).
\end{equation}

Notice that for high frequencies, it is possible that $q_i+q_j>\pi$, and due to the periodic nature of Fourier frequencies, the resulting equivalent mode in the interval $(-\pi,\pi)$ is $q_i+q_j - 2\pi$, which might describe a much lower wavevector, so that a mode that is describing the amplitude of a low length-scale feature will contribute to the a mode describing a large length-scale. This artifact, which results from discretizing space, is known as aliasing, and the need to mitigate aliasing has been known in the context of spectral integration for decades \cite{orszag1971elimination,orszag1972comparison,patterson1971spectral}.

There are several methods to avoid these artifacts, all under the common name of de-aliasing \cite{boyd2001chebyshev}. A simple one, which is implemented here, is creating a de-aliased version of the fields that form part of nonlinearities, by setting to $0$ the amplitude of their Fourier modes that create aliasing artifacts, and using these de-aliased fields to compute nonlinearities.

More specifically, if a field is part of a nonlinearity of order $n$, we find the  values of $q$ such that $nq > 2\pi/\Delta x - q$, and set those modes to $0$. This must also be done for negative frequencies, that is, for every $q<0$ such that $nq < -2\pi/\Delta x+q$. This condition trivially becomes $|q| > 2\pi/[(n+1)\Delta x]$. For instance, for a quadratic nonlinearity, where $n=2$, we set every mode of a field to $0$ for every frequency $q$ such that $|q|>2\pi/(3\Delta x)$. This becomes the commonly known two thirds rule for de-aliasing quadratic nonlinearities. 

In the rest of the paper, it is assumed that nonlinear terms in any equation are written in terms of de-aliased fields, for instance, Eq. \ref{eq:cubicexample} should be rewritten as
\begin{equation}\label{eq:cubicexampledealiased}
    \partial_t\tilde\phi(\bq,t) = \mathcal F\left[\mathcal F^{-1}\left[\tilde\phi^{d_3}(\bq,t)\right]^3\right],
\end{equation}
where the field has been de-aliased before being transformed to real space, such that
\begin{equation}
    \tilde\phi^{d_3}(\bq,t) = \begin{cases} 
      \tilde\phi(\bq,t) & \text{if } |q| < \frac{2\pi}{4\Delta x} \\
      0 & \text{if } |q| > \frac{2\pi}{4\Delta x}
   \end{cases}
\end{equation}

The rest of this paper describes a software package that allows to numerically integrate systems of equations described generally by Eq. \ref{eq:base}. This package is available at \ms{https://github.com/fcaballerop/cuPSS}, and runs natively on GPUs, allowing for fast integration when compared to other packages that run only on CPUs. This package allows equations to be written in a quasi-natural syntax, which is its main strength, and allows fast model prototyping; for instance, a diffusion equation in Fourier space $\partial_t\phi(\bq,t) = -Dq^2\phi(\bq,t)$ can be coded as \ms{dt phi = -D*q\^{}2*phi}. This together with its GPU-accelerated capabilities is what I hope will make this useful to the scientific community.

The next sections describe, respectively, what the package will calculate when given a set of equations, how it does it technically, how to write models, other features such as callback functions, and finally some benchmarking done against other similar packages.

\section{What cuPSS calculates}
cuPSS integrates in time a system of first-order stochastic PDEs for a set of fields $\lbrace\phi_i(\bx,t)\rbrace$ defined on a discrete lattice. Thus, each field follows an equation.
\begin{equation}\label{eq:basereal}
    \partial_t \phi_i = \mathcal L_i \phi_i + \sum_j\mathcal{N}_{ij}[\lbrace\phi_k\rbrace] + \sqrt{2\tilde{D}}\eta_i
\end{equation}
where $\mathcal L_i$ is a linear operator, and $\mathcal{N}_{ij}[\lbrace\phi_k\rbrace]$ is a set of terms, linear or nonlinear on the full set of fields. The nonlinearities are taken to be polynomials of the fields, each of which can be acted upon by a differential operator, so that $\mathcal N_{ij}[\lbrace\phi_k\rbrace]$  can be written as 
\begin{equation}
    \mathcal N_{ij}[\lbrace\phi_j\rbrace] = M_{ij}\prod_{k} \psi_k^{n_k}.
\end{equation}
where $M_{ij}$ are differential operators, and the fields $\psi_k$ are functions of the fields $\phi_i$ operated upon by arbitrary differential operators. This form of writing the nonlinearities, although cumbersome, will facilitate solving these equations pseudo-spectrally.

This form of writing an equation can be illustrated with an example. Consider the KPZ equation \cite{kardar1986dynamic}, describing nonlinear growth of interfaces,
\begin{equation}
\partial_t\phi = \sigma\nabla^2\phi +\lambda(\nabla\phi)^2 + \sqrt{2D}\eta.
\end{equation}

In this case, the only nonlinearity is $\mathcal N = \lambda \psi^2$, where the only field $\psi = \nabla\phi$ is the field we are solving for acted on by the operator $\nabla$.

In order to solve Eq. \ref{eq:basereal} in a pseudo-spectral manner, first it must be written in Fourier space, 
\begin{equation}\label{eq:basefourier}
    \partial_t\tilde\phi_i = \tilde{\mathcal L}_i\tilde\phi_i + \sum_j\tilde{\mathcal{N}}_{ij}[\lbrace\phi_k\rbrace] + \tilde\eta_i,
\end{equation}
where
\begin{equation}
\tilde\phi(\bq,t) = (2\pi)^{-d/2}\int d\bx e^{-i \bq \cdot \bx}\phi(\bx,t), 
\end{equation}
and
where the nonlinearities are written as the general form of Eq. \ref{eq:cubicexample}
\begin{equation}\label{eq:nonlin_general}
    \tilde{\mathcal{N}}_{ij}[\lbrace\phi_j\rbrace] = \tilde M_{ij}\mathcal F\left[\prod_k \mathcal F^{-1}[\tilde\psi_k]^{n_k}.\right]
\end{equation}

The amplitude of the noise has been absorbed in the definition of $\tilde\eta_i$, so that, in Fourier space, the noise has the following variance
\begin{equation}\label{eq:noise_eq}
    \langle \eta(\bq,t)\eta(\bq',t'\rangle = 2\tilde D(\bq)(2\pi)^d\delta^{(d)}(\bq+\bq')\delta(t-t').
\end{equation}

cuPSS will discretize each field $\phi$ in space, in a lattice with spacing $\Delta x$ (and $\Delta y,\Delta z$ if the system is 2-dimensional or 3-dimensional). Given all fields at a time $t$, time is discretized with a timestep size of $\Delta t$, and the next timestep is calculated in a simple finite difference scheme. For simplicity and stability, cuPSS uses an implicit form of the Euler-Maruyama method, where the linear terms are treated in a strong implicit way, equivalent to a Taylor method \cite{kloeden1992numerical}. Usually, one would discretize Eq. \ref{eq:base} in an implicit way by introducing a parameter $\alpha$ such that the discretization in time looks as follows
\begin{equation}
\begin{aligned}
    \frac{\phi(t+\Delta t)-\phi(t)}{\Delta t} =&\;\; \alpha F[\phi(t+\Delta t)]\\&+(1-\alpha)F[\phi(t)] \\ &+ \frac{\sqrt{2D}}{\Delta t}\Delta W_t,
    \end{aligned}
\end{equation}
where $\alpha$ measures the amount of ``implicitness'', and $\Delta W_t$ is a Wiener process with variance $\Delta t$. Choosing the Ito representation of the initial stochastic process allows produces a more natural time discretization of the noise term, so that, in practice, $\Delta W_t$ is calculated as a set of independent noise terms every timestep \cite{kloeden1992numerical}.

Setting $\alpha=1$ corresponds to a fully implicit method for the deterministic part of the equation, sometimes called Milstein scheme \cite{kloeden1992numerical,mannella1997numerical,mannella2000gentle}, while $\alpha=0$ corresponds to the Euler-Maruyama scheme. The Milstein scheme involves having to solve an algebraic equation, which would give greater numerical stability at the sacrifice of speed. What cuPSS does ---which is used in some other widespread packages for the numerical integration of PDEs \footnote{See for example the CFD package OpenFOAM, which allows the user to describe each term in a dynamic equation as explicit or implicit: https://www.openfoam.org}--- is to use a Milstein scheme for the linear terms of each field only, for which the algebraic equation to solve is trivial, and an Euler-Maruyama step for terms that are either linear on different fields, or nonlinear. This offers a way that sacrifices some numerical stability, and might require smaller timestepping, but offers a higher speed of integration. This means that a single time update step for a field $\phi_i$ is
\begin{equation}\label{eq:timeupdate}
    \phi_i(t+\Delta t) = \frac{\phi_i(t) + \Delta t\sum_j \mathcal N_{ij} +\sqrt{\Delta t}\tilde\eta_{i}}{1-\Delta t\tilde{\mathcal{L}}_i}.
\end{equation}

%Notice that although we have defined $\mathcal N_{ij}$ as nonlinearities, in practice the user can write any kind of term as a term to be solved explicitly, including linear terms.

cuPSS will, every timestep, perform the following steps
\begin{enumerate}
    \item Calculate all intermediate terms of the nonlinearities $\psi_k$.
    \item de-alias all intermediate terms $\tilde\psi_k$ according to the highest order nonlinearity in which they are found.
    \item Convert all intermediate terms to real space, and calculate each $\mathcal N_{ij}$ by multiplying its factors.
    \item Convert the nonlinearities to Fourier space.
    \item If any of the fields is stochastic, generate a Gaussian white noise field and apply its corresponding variance $\tilde D$.
    \item Advance each field one timestep according to Eq. \ref{eq:timeupdate}
\end{enumerate}

cuPSS will perform these steps using a GPU if set up to do so, in which case it will use CUDA and some of its packages to considerably speed up the process. cuFFT is used to perform all Fourier transforms, and cuRAND is used to calculate noise realizations if any of the fields are stochastic. If the solver is set to run on CPU, then fftw3 will be used to calculate Fourier transforms, and the standard C++ random library will be used to calculate noise terms. 

%Technically, a cuPSS solver contains a vector of ``field'' objects, which correspond to either a $\phi$ or a $\psi$ field. Each field object contains a vector of ``terms'', which is a set of prefactors (differential operators) and fields, so that the term is defined to be the sum of the prefactors applied to the product of the fields. Each field is then updated as the sum of its terms. Each field and term has arrays that contain its values in both real and Fourier space, as well as a de-aliased version. Every array is duplicated in GPU memory. 

%In the case of GPU acceleration, some efficiency is gained by not going through CPU memory at all. All the arrays are stored in GPU memory, and copied to CPU only when writing data out to disk. This means every update can be a series of CUDA kernel calls without any memory copying. This is done by fields storing pointers to every array that corresponds to every term, as well as arrays containing pointers to every array. This allows the kernels themselves to loop through every term and apply it to its corresponding field without giving control back to the CPU, and without the evolver knowing a priori how many fields and terms it is calculating. See \ms{src/field\_kernels.cu} and \ms{src/term\_kernels.cu} for details.

\section{Writing models}

cuPSS offers functions to make it possible to write models in close to natural language. The steps to integrating a model in cuPSS are (i) creating a system, which holds information about the system size, space and time discretization and data output; (ii) defining the fields that make up the system and the equations that describe their time evolution, and whatever parameters they depend on; (iii) setting an initial condition and callback functions (if any); (iv) evolving the system in time.

These steps, for a single field and a single equation, can be as simple as
\begin{lstlisting}[language=C++]
system.createField("phi", true);
system.addParameter("D", 1.0);
system.addEquation("dt phi + D*q^2*phi = 0");
system.addNoise("phi", "D");
\end{lstlisting}
where it is hopefully natural enough to read the equation line as a noisy diffusion equation for a field $\phi$, with diffusion constant $D$, i.e. $\partial_t\tilde\phi + Dq^2\phi=0$.

Writing equations is done through a parser that follows rules similar to other available numerical solvers; when writing an equation for a dynamic field, named for instance \ms{phi}, the string corresponding to it must look as follows

\ms{dt phi + lhs = rhs}

where \ms{lhs} represents terms that are treated implicitly in time, while \ms{rhs} represents terms that are treated explicitly, as described above. If there are no implicit terms, the string becomes \ms{dt phi = rhs}; and if there are only implicit terms, it becomes \ms{dt phi + lhs = 0}.

Implicit terms must be linear in the field for which we are writing an equation, so each term must be \ms{op * phi}, where \ms{op} can be a combination of numbers or differential operators that form its prefactor.

Explicit terms can be any combination of differential operators and fields. It is important to have in mind that, since the equation is written in Fourier space, differential operators will be applied on the whole term, and not on any of the factors. For instance, if we write \ms{dt phi = -q\^{}2*phi*psi}, where \ms{psi} is some other field, this equation correspondg to $\partial_t\phi = \nabla^2(\phi\psi)$.

If we wanted the Laplacian to act on a single field, $\partial_t\phi = \psi\nabla^2\phi$, we would need to define an intermediate field for $\nabla^2\phi$, cuPSS will then appropriately de-alias both this field and $\psi$, and compute their product, we would then need two equations:

\ms{dt phi = psi*lapphi}

\ms{lapphi = -q\^{}2*phi}

Equations for intermediate field $\psi_i$ are written similarly, except there is no \ms{dt} operator. Again, the left hand side must only have terms linear in the field the equation is describing, and terms in the right hand side can be general products of the rest of the fields. For instance, we can write, for a field \ms{psi}

\ms{psi + psi * q\^{}2 = -q\^{}2phi}

which is equivalent to a field $\psi$ defined as a function of $\phi$, such that $\tilde\psi(q,t)+q^2\tilde\psi(q,t) = -q^2\tilde\phi(q,t)$, or alternatively, $\tilde\psi(q,t) = -q^2(1+q^2)^{-1}\tilde\phi(q,t)$.

\subsection*{Reserved keywords}
The equation parser has certain keywords that must not be used as parameter or field names. These keywords are the following

\ms{dt}. This keyword should be read as $\partial_t$, and indicates we are writing the equation of motion for a field. It should be the first two characters of a dynamical equation, and the parser will interpret that whatever comes after it it the name of the field we are writing an equation for, thus if the parser is given \ms{dt phi ...}, it will interpret that string as the dynamic equation for a field with the name \ms{phi}.

\ms{q}. This keyword is interpreted as a derivative operator, it must be accompanied by a caret and an even number, and should be interpreted as powers of the laplacian, for example \ms{q\^{}2} should be read as a $-\nabla^2$ operator. 

\ms{iqx}, \ms{iqy} and \ms{iqz}. These represent derivatives in the $x$, $y$ and $z$ directions respectively. For example \ms{dt phi = iqx\^{}2*phi} will be interpreted as the equation $\partial_t\phi = \partial_x^2\phi$. Notice there must be an asterisk between every pair of factors.

\ms{1/q}. This keyword represents a division over the absolute value of the frequency in Fourier space. It can be useful to write certain nonlocal kernels \cite{bray2001interface,besse2023interface}.

\subsection{Adding noise}

Stochasticity can be added to a field through the function \ms{evolver::addNoise}, which will turn a field's internal boolean \ms{isNoisy} to true and add the noise's appropriate prefactor describing $\tilde D$, so that noise correlations are written as in Eq. \ref{eq:noise_eq}. This prefactor can be a product of a real number with a power of \ms{q\^{}2} and \ms{1/q}. For instance, the noise corresponding to a mass conserving continuity equation, with $\tilde D = D q^2$ would be added to a field \ms{phi} with these two lines
\begin{lstlisting}[language=C++]
system.addParameter("D", 0.01);
system.addNoise("phi", "D*q^2");
\end{lstlisting}

\section{Example}

Since spectral methods usually overperform real space methods when integrating equations with high order gradient terms, I provide a simple example for one such model. Commonly known in the literature as models B or Cahn-Hilliard model, this equation describes binary mixtures and their transition between uniform and phase separated regimes \cite{hohenberg1977theory}. A phase field $\phi(r,t)$ represents local relative concentration of the two species of the binary mixture, and follows a continuity equation $\partial_t\phi + \nabla\cdot J=0$, where the current follows the gradient of a chemical potential $J=-\nabla\mu$. The chemical potential is derived from a free energy $\mathcal G[\phi]$, thus $\mu = \delta\mathcal G/\delta\phi$, which is taken to be a Landau-Ginzburg expansion on the field $\phi$ and its gradients.

\begin{equation}
    \mathcal G[\phi] = \int d^2r \frac{a}{2}\phi^2 + \frac{b}{4}\phi^4 + \frac{k}{2}(\nabla\phi)^2
\end{equation}

The full equation of motion for $\tilde\phi$ can thus be written, in this case ignoring stochasticity, as
\begin{equation}
    \partial_t\tilde\phi + q^2(a + k q^2)\tilde\phi = -b q^2 \mathcal F\left[(\mathcal F^{-1}[\tilde\phi])^3\right].
\end{equation}

\begin{figure}
    \includegraphics[width=0.48\textwidth]{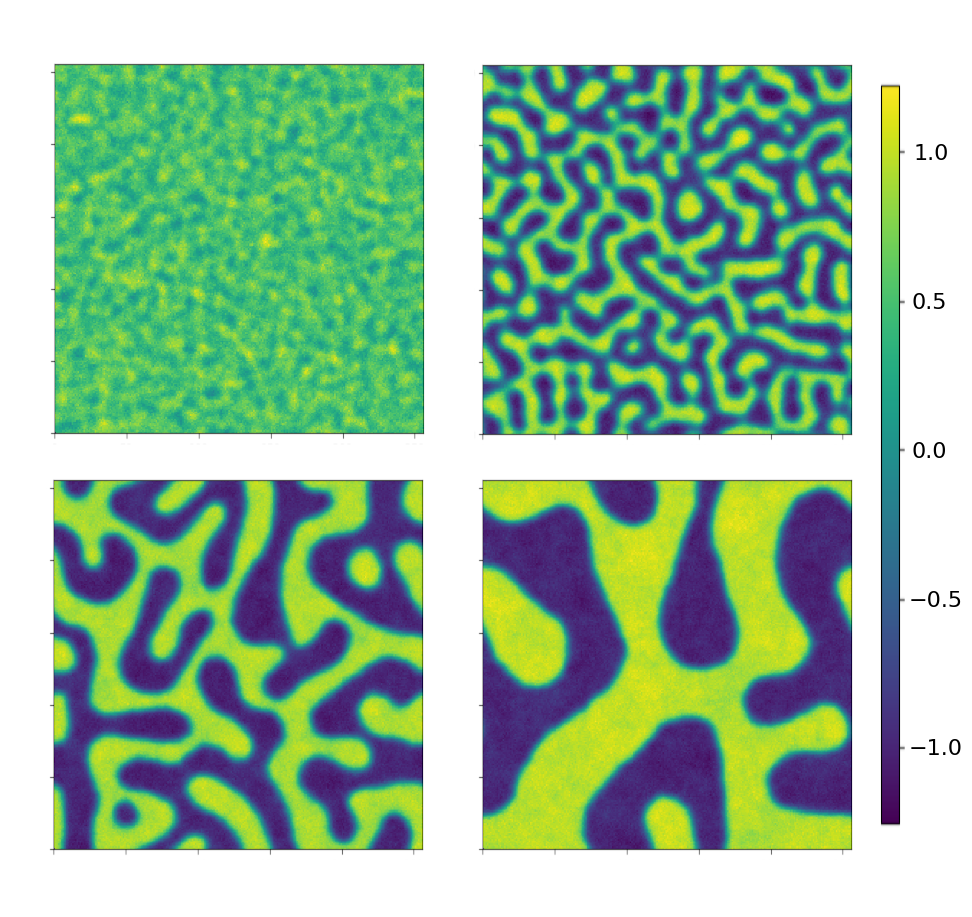}
    \caption{Snapshots at different times of the phase field $\phi$ obeying Cahn-Hilliard dynamics, as calculated by the example solver of the main text, showing standard spinodal decomposition. The parameters are $-a=b=1,k=4$ for a system of size $256\times 256$, a timestep of $\Delta t = 0.1$. The initial condition is a small random perturbation over a uniform state, and the snapshots are taken at times $t=10, 100, 1000, 5000$.}
    \label{fig:modelbsnapshots}
\end{figure}

A solver for this equation involves three parameters, $a$, $b$ and $k$, and a single field $\phi$. The following short code snippet creates such a system, in this case discretized in a two-dimensional lattice of size \ms{Nx}$\times$\ms{Ny}, with lattice sites of size $\Delta x=\Delta y$, and with a timestep $\Delta t$:

\begin{lstlisting}[language=C++]
evolver system(1, Nx, Ny, dx, dy, dt, output);

system.addParameter("a", -1.0);
system.addParameter("b", 1.0);
system.addParameter("k", 4.0);

system.createField("phi", 1);

system.addEquation(
"dt phi + (a*q^2 + k*q^4)*phi = -b*q^2*phi^3");
\end{lstlisting}

Since the chosen parameters are $-a=b=1$ and $k=4$, if the system is initialized with a small noise, it should display an initial instability that leads the system into a spinodal decomposition into two regions with $\phi = \pm 1$, separated by interfaces of thickness $\ell\approx \sqrt{-k/a}=2$. Fig. \ref{fig:modelbsnapshots} shows some snapshots of the output created by the solver above.

\section{Callback functions}

cuPSS offers direct dynamic access to all fields and terms, meaning the user can access directly the value of all fields during the progress of a simulation. It offers an easy way to do so through callback functions; functions that are called giving access to the real values of a field at the end of a timestep. There are many applications for callback functions, such as applying boundary conditions. Since a Fourier spectral decomposition assumes periodic boundary conditions, applying other boundary conditions through callback functions should be done with caution. Setting Dirichlet boundary conditions, which would be properly done spectrally through the choice of a different spectral basis \cite{boyd2001chebyshev}, is also usually done by setting a buffer zone of infinite stiffness in the boundaries of the lattice \cite{zhao2024asymmetric,risthaus2024imposing}, i.e. by setting a field to a certain value at a range of positions close to the boundary, making sure this range is bigger than all lengthscales of the problem. 

We illustrate the issue of setting boundary conditions this way by solving a heat diffusion equation in one dimension between two heat baths at two different temperatures $T_1$ and $T_2$. The temperature $T(x,t)$ obeys $\partial_t T(x,t) = \partial_x^2T(x,t)$, with the boundary conditions $T(0,t) = T_1$ and $T(L,t) = T_2$. The steady state solution is $T(x,t) = T_1 + (T_2-T_1)x/L$. To implement this boundary condition we first need to define a callback function which must take five arguments, a pointer to the evolver itself, giving access to the full system, a pointer to the array storing the real space value of the field we are setting a boundary for, and the three dimensional size of the system (the last two parameters will be $1$ for a one-dimensional system)
\begin{lstlisting}[language=C++]
#define T1 10.0
#define T2 0.0
void boundary_diffusion(evolver *sys, float2 * array, int Nx, int Ny, int Nz)
{
    array[0] = T1;
    array[Nx-1] = T2;
}
\end{lstlisting}

We then assign the boundary condition, by setting the \ms{hasCB} flag on that field to \ms{true}, and assigning a pointer to the callback function to that field's \ms{callback} component
\begin{lstlisting}[language=C++]
system.fieldsMap["phi"]->hasCB = true;
system.fieldsMap["phi"]->callback = boundary_diffusion;
\end{lstlisting}

\begin{figure}
    \centering
    \includegraphics[width=0.45\textwidth]{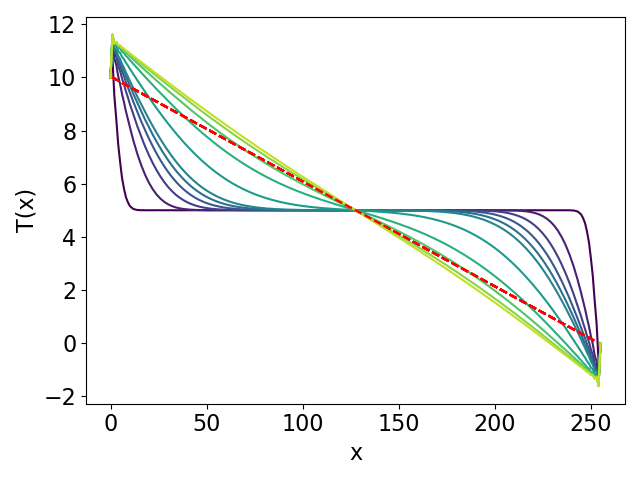}

    \includegraphics[width=0.45\textwidth]{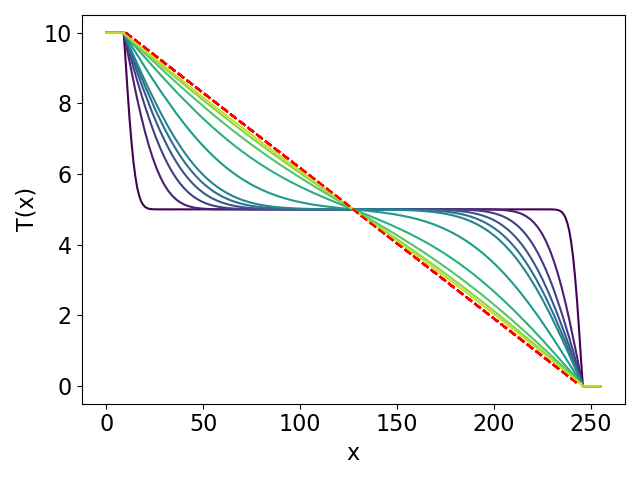}
    \caption{Solution of a diffusion heat diffusion equation between two heat baths at two different temperatures in discretized space with $256$ lattice sites. The dashed red line is the analytical solution, while the other lines represent the numerical solution starting from a constant value $T(x,0) = 5$, and where lighter colors represent later times. The plot at the top show the solution where the heat baths are applied by setting $T(0,t)=10$ and $T(256,t)=0$ every timestep, showing the numerical artifact that results from setting Dirichlet boundary conditions on an intrinsically periodic field, in which the numerical steady state deviates from the analytical solution. The plot in the bottom shows the same, where the heat baths are applied in a frame of width $10$ around the boundary, i.e. $T(x<10,t)=10$ and $T(x>246,t)=0$. This eliminates the numerical artifact, allowing for smoother Fourier representations of the field at any time, and obtaining a solution that now converges to the analytical solution.}
    \label{fig:diff1}
\end{figure}

Running a diffusion solver with the previous boundary condition would result in a numerical artifact coming from creating a high gradient at $x=0=Nx$, created by the periodic nature of the problem, as seen in Fig \ref{fig:diff1}. If instead we create a frame, thus setting the boundaries in a small region around the boundaries, we get rids of the numerical artifacts. Specifically, if we changed the boundary condition to 
\begin{lstlisting}[language=C++]
#define T1 10.0
#define T2 0.0
void boundary_diffusion(evolver *sys, float2 * array, int Nx, int Ny, int Nz)
{
    for (int i = 0; i < 10; i++)
    {
        array[i] = T1;
        array[Nx-1-i] = T2;
    }
}
\end{lstlisting}
we would obtain a solution that now converges to the actual analytical solution for the steady state $T(x, t\rightarrow\infty)$, which should be a straight line between the temperatures of both heat baths, as shown also in Fig \ref{fig:diff1}.

\section{Benchmarking}

This section shows limited benchmarking of cuPSS in different systems, a consumer level system with a Intel Core i7-13700K CPU and a NVIDIA RTX 3060 Ti GPU, and a high-performance computing cluster, running in a few different GPUs. For reference, some benchmarks are compared to the same systems integrated in two different packages, FEniCSx \cite{barrata2023dolfinx}, a finite difference solver for PDEs written in weak form, and Dedalus \cite{burns2020dedalus}, another spectral solver that implements the same spectral techniques as cuPSS, together with extra features such as non-flat spaces and other spectral basis. This section shows the main advantage of cuPSS, which achieves much higher speeds compared to other packages through the direct implementation of the library on CUDA C++, and having a simplified codebase that relies more on speed than depth of features and breadth.

The benchmarks have been run with two example solvers, for models B and H \cite{hohenberg1977theory}, as these two solvers implement both a simple example with a single field and nonlinear term, and a more complex example with several fields and constraints. The model B solver is as described in the Example section, while the model H solver is supplied in the package's repository. The benchmark results are given in terms of average time taken to integrate one timestep, calculated as an average over a number of timesteps that ranges from 100 to 10000 depending on the problem and system size. If there is any overhead in any of these solvers, associated to preparing the problem, reserving memory and so on, this time has not been taken into account in the time measurements.

\begin{figure}
    \centering
    \includegraphics[width=0.5\textwidth]{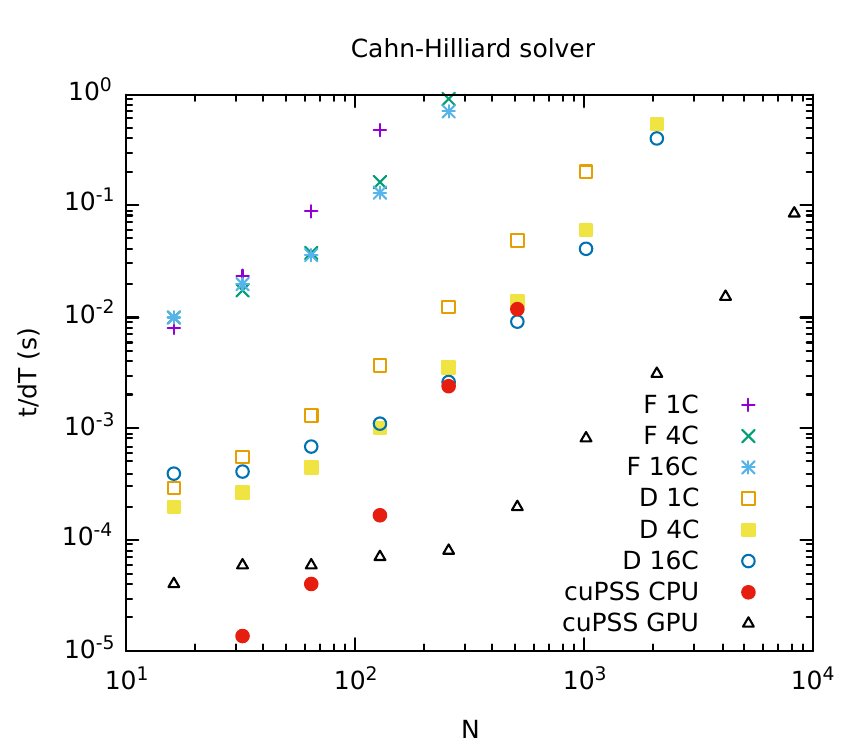}
    \caption{Time $t$ it takes on average to integrate one timestep $dT$, in seconds, as a function of the size of a 2-dimensional system, $N$ being its side length. Each of these runs is done over 10.000 timesteps in a system of size $N\times N$. cuPSS in this case is running on a single core on an i7-13700K, or on a 3060 Ti in the case of the GPU runs. In the legend, F and D stand for FEniCSx and Dedalus respectively, and the number is the number of cores they run on, on the same system.}
    \label{fig:modelb_local}
\end{figure}

\begin{figure}
    \centering
    \includegraphics[width=0.5\textwidth]{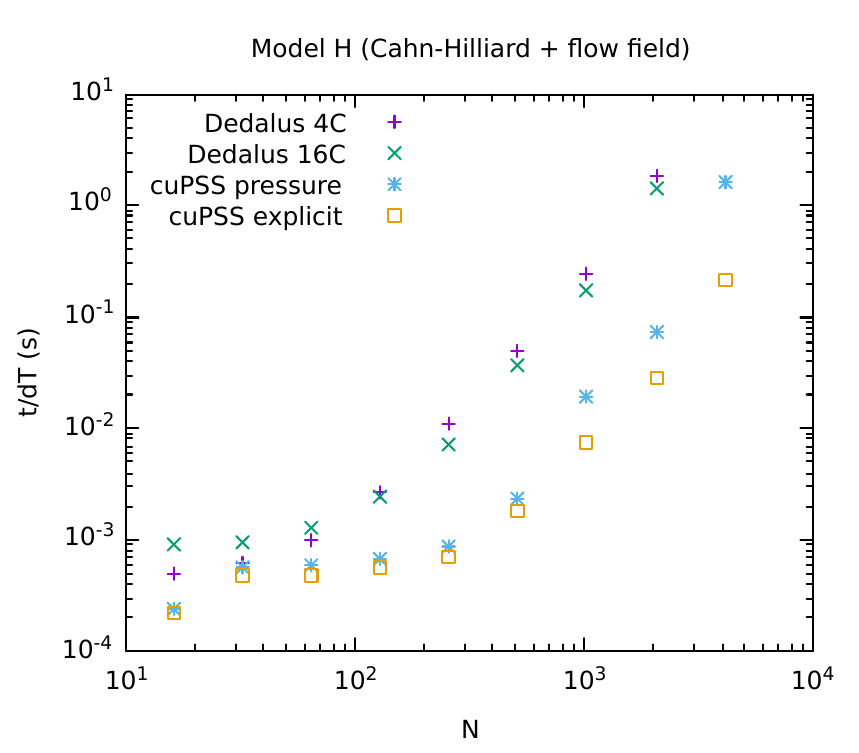}
    \caption{Same as Fig. \ref{fig:modelb_local}, but for a solver for model H, considerably more complicated and with several intermediate fields to calculate. In the case of cuPSS, the pressure and explicit refer to solvers that solve the Poisson equation for the pressure and the flow directly.}
    \label{fig:modelh_local}
\end{figure}

\begin{figure}
    \centering
    \includegraphics[width=0.5\textwidth]{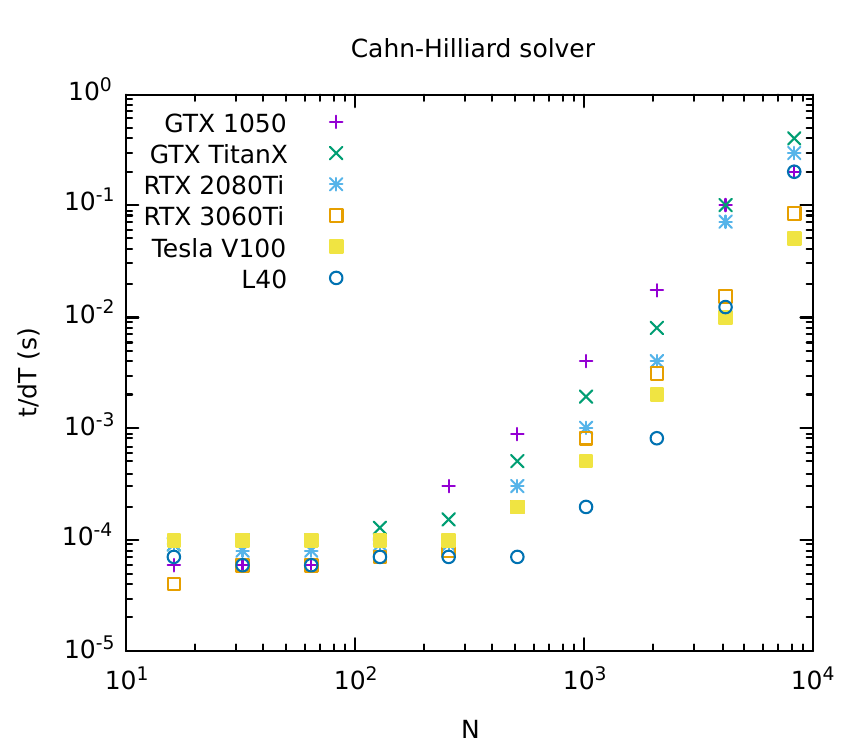}
    \caption{Comparison of speed in several different graphics cards for a Cahn-Hilliard solver on cuPSS. There is no appreciable difference in speed until the system size is large enough, at which point the faster graphics cards are able to process the bigger textures associated to each field in less time.}
    \label{fig:modelb_hpcc}
\end{figure}

Figure \ref{fig:modelb_local} shows the time per timestep to integrate the Cahn-Hilliard dynamics, compared between different solvers on different configurations, Figure \ref{fig:modelh_local} shows the same for a solver of model H, and finally Figure \ref{fig:modelb_hpcc} shows again Cahn-Hilliard dynamics benchmarks in four different types of GPUs.

There is one series of benchmarking done on CPU, while all others run on GPUs. CPUs are only more efficient when integrating small systems, in this case when the system size is less than $128\times128$. Due to the limited extent of this benchmarking, it is good practive to test any particular solver on different CPUs and GPUs to verify which option offers a higher speed. Benchmarks have also been run only on system sizes that are powers of 2. These system sizes result in the best speeds given the nature of both the way CUDA launches kernels on GPUs, and the way FFT algorithms work. This means running on system sizes different than powers of 2 might show a loss of speed compared to the system size. In the worst case scenario, a system size different than $2^n$ might run at the same speed as a system size that is equal to the next power of $2$, specially when running on GPUs.

The main result of this section is that cuPSS offers a very significant improvement in speed, of up to a few orders of magnitude, specially in bigger lattice sizes, when compared to other popular finite-difference solvers. This is thanks to several factors, such as relying on GPUs and sacrificing some breadth of application by restricting systems to Fourier representations, as well as keeping a small codebase that does not introduce unnecessary features for simple applications.

\section{Future development}

cuPSS is offered as open source with several examples, and its source code can be found on github \footnote{\lowercase{h}ttps://github.com/fcaballerop/cuPSS} under an MIT license, and is thus open to other people to further develop or contribute to. Additionally, cuPSS has a relatively small code base of around 3000 lines of code, making it in principle easy for developers to change and modify to meet their needs. This small code base is at the same time one of its biggest strengths when compared to other similar packages, making the code very accessible to new users. There is ample space for cuPSS to grow and incorporate new features, such as higher dimensionality spaces, curved spaces, and so on, without much sacrifice to its speed.

In conclusion, cuPSS offers a quick way to prototype and test continuum theories for a variety of different physical systems. It eliminates the need to know details about particular algorithmic implementations of numerical methods, and provides a simple way to write theories in a quasi-natural language, without deep knowledge of either C++ or CUDA.

With continuum development and optimization, cuPSS has the potential to become a useful tool for researchers in very different areas, as it is already being used in work, specially in its purely deterministic form, some of which has already been published \cite{caballero2023vorticity,gulati2024traveling,zhao2024asymmetric}. Written in the CUDA C++ standard, it should be accessible to most research groups to, not only use, but easily modify and adapt to their needs.

\begin{acknowledgments}
I thank Paarth Gulati for help testing during the early parts of the development of cuPSS, and for suggesting its name, and Daniel Hellstein for guidance in the development of some of the solver components. This work was done during stays at the University of California, Santa Barbara and Brandeis University, and was partially funded by NSF grants DMR-2041459 \& DMR-2011846. This work was also supported by the NSF through DMR 2309635 and the Brandeis Center for Bioinspired Soft Materials, an NSF MRSEC (DMR-2011846). We also acknowledge computational support from NSF XSEDE computing resources allocation TG-MCB090163, National Energy Research Scientific Computing Center (NERSC), a Department of Energy Office of Science User Facility using NERSC award BES-ERCAP0026774, and the Brandeis HPCC which is partially supported by the NSF through DMR-MRSEC 2011846 and OAC-1920147.
\end{acknowledgments}

\medskip

\bibliography{refs}

%apsrev4-2.bst 2019-01-14 (MD) hand-edited version of apsrev4-1.bst
%Control: key (0)
%Control: author (8) initials jnrlst
%Control: editor formatted (1) identically to author
%Control: production of article title (0) allowed
%Control: page (0) single
%Control: year (1) truncated
%Control: production of eprint (0) enabled
\begin{thebibliography}{35}%
\makeatletter
\providecommand \@ifxundefined [1]{%
 \@ifx{#1\undefined}
}%
\providecommand \@ifnum [1]{%
 \ifnum #1\expandafter \@firstoftwo
 \else \expandafter \@secondoftwo
 \fi
}%
\providecommand \@ifx [1]{%
 \ifx #1\expandafter \@firstoftwo
 \else \expandafter \@secondoftwo
 \fi
}%
\providecommand \natexlab [1]{#1}%
\providecommand \enquote  [1]{``#1''}%
\providecommand \bibnamefont  [1]{#1}%
\providecommand \bibfnamefont [1]{#1}%
\providecommand \citenamefont [1]{#1}%
\providecommand \href@noop [0]{\@secondoftwo}%
\providecommand \href [0]{\begingroup \@sanitize@url \@href}%
\providecommand \@href[1]{\@@startlink{#1}\@@href}%
\providecommand \@@href[1]{\endgroup#1\@@endlink}%
\providecommand \@sanitize@url [0]{\catcode `\\12\catcode `\$12\catcode `\&12\catcode `\#12\catcode `\^12\catcode `\_12\catcode `\%12\relax}%
\providecommand \@@startlink[1]{}%
\providecommand \@@endlink[0]{}%
\providecommand \url  [0]{\begingroup\@sanitize@url \@url }%
\providecommand \@url [1]{\endgroup\@href {#1}{\urlprefix }}%
\providecommand \urlprefix  [0]{URL }%
\providecommand \Eprint [0]{\href }%
\providecommand \doibase [0]{https://doi.org/}%
\providecommand \selectlanguage [0]{\@gobble}%
\providecommand \bibinfo  [0]{\@secondoftwo}%
\providecommand \bibfield  [0]{\@secondoftwo}%
\providecommand \translation [1]{[#1]}%
\providecommand \BibitemOpen [0]{}%
\providecommand \bibitemStop [0]{}%
\providecommand \bibitemNoStop [0]{.\EOS\space}%
\providecommand \EOS [0]{\spacefactor3000\relax}%
\providecommand \BibitemShut  [1]{\csname bibitem#1\endcsname}%
\let\auto@bib@innerbib\@empty
%</preamble>
\bibitem [{\citenamefont {Forster}\ \emph {et~al.}(1977)\citenamefont {Forster}, \citenamefont {Nelson},\ and\ \citenamefont {Stephen}}]{forster1977large}%
  \BibitemOpen
  \bibfield  {author} {\bibinfo {author} {\bibfnamefont {D.}~\bibnamefont {Forster}}, \bibinfo {author} {\bibfnamefont {D.~R.}\ \bibnamefont {Nelson}},\ and\ \bibinfo {author} {\bibfnamefont {M.~J.}\ \bibnamefont {Stephen}},\ }\bibfield  {title} {\bibinfo {title} {Large-distance and long-time properties of a randomly stirred fluid},\ }\href@noop {} {\bibfield  {journal} {\bibinfo  {journal} {Physical Review A}\ }\textbf {\bibinfo {volume} {16}},\ \bibinfo {pages} {732} (\bibinfo {year} {1977})}\BibitemShut {NoStop}%
\bibitem [{\citenamefont {Ma}\ and\ \citenamefont {Mazenko}(1975)}]{ma1975critical}%
  \BibitemOpen
  \bibfield  {author} {\bibinfo {author} {\bibfnamefont {S.-k.}\ \bibnamefont {Ma}}\ and\ \bibinfo {author} {\bibfnamefont {G.~F.}\ \bibnamefont {Mazenko}},\ }\bibfield  {title} {\bibinfo {title} {Critical dynamics of ferromagnets in 6-$\varepsilon$ dimensions: General discussion and detailed calculation},\ }\href@noop {} {\bibfield  {journal} {\bibinfo  {journal} {Physical Review B}\ }\textbf {\bibinfo {volume} {11}},\ \bibinfo {pages} {4077} (\bibinfo {year} {1975})}\BibitemShut {NoStop}%
\bibitem [{\citenamefont {Hohenberg}\ and\ \citenamefont {Halperin}(1977)}]{hohenberg1977theory}%
  \BibitemOpen
  \bibfield  {author} {\bibinfo {author} {\bibfnamefont {P.~C.}\ \bibnamefont {Hohenberg}}\ and\ \bibinfo {author} {\bibfnamefont {B.~I.}\ \bibnamefont {Halperin}},\ }\bibfield  {title} {\bibinfo {title} {Theory of dynamic critical phenomena},\ }\href@noop {} {\bibfield  {journal} {\bibinfo  {journal} {Reviews of Modern Physics}\ }\textbf {\bibinfo {volume} {49}},\ \bibinfo {pages} {435} (\bibinfo {year} {1977})}\BibitemShut {NoStop}%
\bibitem [{\citenamefont {Tjhung}\ \emph {et~al.}(2018)\citenamefont {Tjhung}, \citenamefont {Nardini},\ and\ \citenamefont {Cates}}]{tjhung2018cluster}%
  \BibitemOpen
  \bibfield  {author} {\bibinfo {author} {\bibfnamefont {E.}~\bibnamefont {Tjhung}}, \bibinfo {author} {\bibfnamefont {C.}~\bibnamefont {Nardini}},\ and\ \bibinfo {author} {\bibfnamefont {M.~E.}\ \bibnamefont {Cates}},\ }\bibfield  {title} {\bibinfo {title} {Cluster phases and bubbly phase separation in active fluids: reversal of the ostwald process},\ }\href@noop {} {\bibfield  {journal} {\bibinfo  {journal} {Physical Review X}\ }\textbf {\bibinfo {volume} {8}},\ \bibinfo {pages} {031080} (\bibinfo {year} {2018})}\BibitemShut {NoStop}%
\bibitem [{\citenamefont {Kardar}\ \emph {et~al.}(1986)\citenamefont {Kardar}, \citenamefont {Parisi},\ and\ \citenamefont {Zhang}}]{kardar1986dynamic}%
  \BibitemOpen
  \bibfield  {author} {\bibinfo {author} {\bibfnamefont {M.}~\bibnamefont {Kardar}}, \bibinfo {author} {\bibfnamefont {G.}~\bibnamefont {Parisi}},\ and\ \bibinfo {author} {\bibfnamefont {Y.-C.}\ \bibnamefont {Zhang}},\ }\bibfield  {title} {\bibinfo {title} {Dynamic scaling of growing interfaces},\ }\href@noop {} {\bibfield  {journal} {\bibinfo  {journal} {Physical Review Letters}\ }\textbf {\bibinfo {volume} {56}},\ \bibinfo {pages} {889} (\bibinfo {year} {1986})}\BibitemShut {NoStop}%
\bibitem [{\citenamefont {Bray}\ \emph {et~al.}(2001)\citenamefont {Bray}, \citenamefont {Cavagna},\ and\ \citenamefont {Travasso}}]{bray2001interface}%
  \BibitemOpen
  \bibfield  {author} {\bibinfo {author} {\bibfnamefont {A.~J.}\ \bibnamefont {Bray}}, \bibinfo {author} {\bibfnamefont {A.}~\bibnamefont {Cavagna}},\ and\ \bibinfo {author} {\bibfnamefont {R.~D.}\ \bibnamefont {Travasso}},\ }\bibfield  {title} {\bibinfo {title} {Interface fluctuations, {B}urgers equations, and coarsening under shear},\ }\href@noop {} {\bibfield  {journal} {\bibinfo  {journal} {Physical Review E}\ }\textbf {\bibinfo {volume} {65}},\ \bibinfo {pages} {016104} (\bibinfo {year} {2001})}\BibitemShut {NoStop}%
\bibitem [{\citenamefont {Caballero}\ \emph {et~al.}(2018{\natexlab{a}})\citenamefont {Caballero}, \citenamefont {Nardini}, \citenamefont {van Wijland},\ and\ \citenamefont {Cates}}]{caballero2018strong}%
  \BibitemOpen
  \bibfield  {author} {\bibinfo {author} {\bibfnamefont {F.}~\bibnamefont {Caballero}}, \bibinfo {author} {\bibfnamefont {C.}~\bibnamefont {Nardini}}, \bibinfo {author} {\bibfnamefont {F.}~\bibnamefont {van Wijland}},\ and\ \bibinfo {author} {\bibfnamefont {M.~E.}\ \bibnamefont {Cates}},\ }\bibfield  {title} {\bibinfo {title} {Strong coupling in conserved surface roughening: a new universality class?},\ }\href@noop {} {\bibfield  {journal} {\bibinfo  {journal} {Physical review letters}\ }\textbf {\bibinfo {volume} {121}},\ \bibinfo {pages} {020601} (\bibinfo {year} {2018}{\natexlab{a}})}\BibitemShut {NoStop}%
\bibitem [{\citenamefont {Besse}\ \emph {et~al.}(2023)\citenamefont {Besse}, \citenamefont {Fausti}, \citenamefont {Cates}, \citenamefont {Delamotte},\ and\ \citenamefont {Nardini}}]{besse2023interface}%
  \BibitemOpen
  \bibfield  {author} {\bibinfo {author} {\bibfnamefont {M.}~\bibnamefont {Besse}}, \bibinfo {author} {\bibfnamefont {G.}~\bibnamefont {Fausti}}, \bibinfo {author} {\bibfnamefont {M.~E.}\ \bibnamefont {Cates}}, \bibinfo {author} {\bibfnamefont {B.}~\bibnamefont {Delamotte}},\ and\ \bibinfo {author} {\bibfnamefont {C.}~\bibnamefont {Nardini}},\ }\bibfield  {title} {\bibinfo {title} {Interface roughening in nonequilibrium phase-separated systems},\ }\href@noop {} {\bibfield  {journal} {\bibinfo  {journal} {Physical Review Letters}\ }\textbf {\bibinfo {volume} {130}},\ \bibinfo {pages} {187102} (\bibinfo {year} {2023})}\BibitemShut {NoStop}%
\bibitem [{\citenamefont {Wittkowski}\ \emph {et~al.}(2014)\citenamefont {Wittkowski}, \citenamefont {Tiribocchi}, \citenamefont {Stenhammar}, \citenamefont {Allen}, \citenamefont {Marenduzzo},\ and\ \citenamefont {Cates}}]{wittkowski2014scalar}%
  \BibitemOpen
  \bibfield  {author} {\bibinfo {author} {\bibfnamefont {R.}~\bibnamefont {Wittkowski}}, \bibinfo {author} {\bibfnamefont {A.}~\bibnamefont {Tiribocchi}}, \bibinfo {author} {\bibfnamefont {J.}~\bibnamefont {Stenhammar}}, \bibinfo {author} {\bibfnamefont {R.~J.}\ \bibnamefont {Allen}}, \bibinfo {author} {\bibfnamefont {D.}~\bibnamefont {Marenduzzo}},\ and\ \bibinfo {author} {\bibfnamefont {M.~E.}\ \bibnamefont {Cates}},\ }\bibfield  {title} {\bibinfo {title} {Scalar $\varphi$ 4 field theory for active-particle phase separation},\ }\href@noop {} {\bibfield  {journal} {\bibinfo  {journal} {Nature communications}\ }\textbf {\bibinfo {volume} {5}},\ \bibinfo {pages} {4351} (\bibinfo {year} {2014})}\BibitemShut {NoStop}%
\bibitem [{\citenamefont {Tiribocchi}\ \emph {et~al.}(2015)\citenamefont {Tiribocchi}, \citenamefont {Wittkowski}, \citenamefont {Marenduzzo},\ and\ \citenamefont {Cates}}]{tiribocchi2015active}%
  \BibitemOpen
  \bibfield  {author} {\bibinfo {author} {\bibfnamefont {A.}~\bibnamefont {Tiribocchi}}, \bibinfo {author} {\bibfnamefont {R.}~\bibnamefont {Wittkowski}}, \bibinfo {author} {\bibfnamefont {D.}~\bibnamefont {Marenduzzo}},\ and\ \bibinfo {author} {\bibfnamefont {M.~E.}\ \bibnamefont {Cates}},\ }\bibfield  {title} {\bibinfo {title} {Active model h: scalar active matter in a momentum-conserving fluid},\ }\href@noop {} {\bibfield  {journal} {\bibinfo  {journal} {Physical review letters}\ }\textbf {\bibinfo {volume} {115}},\ \bibinfo {pages} {188302} (\bibinfo {year} {2015})}\BibitemShut {NoStop}%
\bibitem [{\citenamefont {Caballero}\ \emph {et~al.}(2018{\natexlab{b}})\citenamefont {Caballero}, \citenamefont {Nardini},\ and\ \citenamefont {Cates}}]{caballero2018bulk}%
  \BibitemOpen
  \bibfield  {author} {\bibinfo {author} {\bibfnamefont {F.}~\bibnamefont {Caballero}}, \bibinfo {author} {\bibfnamefont {C.}~\bibnamefont {Nardini}},\ and\ \bibinfo {author} {\bibfnamefont {M.~E.}\ \bibnamefont {Cates}},\ }\bibfield  {title} {\bibinfo {title} {From bulk to microphase separation in scalar active matter: a perturbative renormalization group analysis},\ }\href@noop {} {\bibfield  {journal} {\bibinfo  {journal} {Journal of Statistical Mechanics: Theory and Experiment}\ }\textbf {\bibinfo {volume} {2018}},\ \bibinfo {pages} {123208} (\bibinfo {year} {2018}{\natexlab{b}})}\BibitemShut {NoStop}%
\bibitem [{\citenamefont {Caballero}\ and\ \citenamefont {Marchetti}(2022)}]{caballero2022activity}%
  \BibitemOpen
  \bibfield  {author} {\bibinfo {author} {\bibfnamefont {F.}~\bibnamefont {Caballero}}\ and\ \bibinfo {author} {\bibfnamefont {M.~C.}\ \bibnamefont {Marchetti}},\ }\bibfield  {title} {\bibinfo {title} {Activity-suppressed phase separation},\ }\href@noop {} {\bibfield  {journal} {\bibinfo  {journal} {Physical Review Letters}\ }\textbf {\bibinfo {volume} {129}},\ \bibinfo {pages} {268002} (\bibinfo {year} {2022})}\BibitemShut {NoStop}%
\bibitem [{\citenamefont {Chat{\'e}}\ and\ \citenamefont {Solon}(2024)}]{chate2024dynamic}%
  \BibitemOpen
  \bibfield  {author} {\bibinfo {author} {\bibfnamefont {H.}~\bibnamefont {Chat{\'e}}}\ and\ \bibinfo {author} {\bibfnamefont {A.}~\bibnamefont {Solon}},\ }\bibfield  {title} {\bibinfo {title} {Dynamic scaling of two-dimensional polar flocks},\ }\href@noop {} {\bibfield  {journal} {\bibinfo  {journal} {arXiv preprint arXiv:2403.03804}\ } (\bibinfo {year} {2024})}\BibitemShut {NoStop}%
\bibitem [{\citenamefont {Besse}\ \emph {et~al.}(2022)\citenamefont {Besse}, \citenamefont {Chat{\'e}},\ and\ \citenamefont {Solon}}]{besse2022metastability}%
  \BibitemOpen
  \bibfield  {author} {\bibinfo {author} {\bibfnamefont {M.}~\bibnamefont {Besse}}, \bibinfo {author} {\bibfnamefont {H.}~\bibnamefont {Chat{\'e}}},\ and\ \bibinfo {author} {\bibfnamefont {A.}~\bibnamefont {Solon}},\ }\bibfield  {title} {\bibinfo {title} {Metastability of constant-density flocks},\ }\href@noop {} {\bibfield  {journal} {\bibinfo  {journal} {Physical Review Letters}\ }\textbf {\bibinfo {volume} {129}},\ \bibinfo {pages} {268003} (\bibinfo {year} {2022})}\BibitemShut {NoStop}%
\bibitem [{\citenamefont {Lande}\ \emph {et~al.}(2003)\citenamefont {Lande}, \citenamefont {Engen},\ and\ \citenamefont {Saether}}]{lande2003stochastic}%
  \BibitemOpen
  \bibfield  {author} {\bibinfo {author} {\bibfnamefont {R.}~\bibnamefont {Lande}}, \bibinfo {author} {\bibfnamefont {S.}~\bibnamefont {Engen}},\ and\ \bibinfo {author} {\bibfnamefont {B.-E.}\ \bibnamefont {Saether}},\ }\href@noop {} {\emph {\bibinfo {title} {Stochastic population dynamics in ecology and conservation}}}\ (\bibinfo  {publisher} {Oxford University Press, USA},\ \bibinfo {year} {2003})\BibitemShut {NoStop}%
\bibitem [{\citenamefont {Wilkinson}(2018)}]{wilkinson2018stochastic}%
  \BibitemOpen
  \bibfield  {author} {\bibinfo {author} {\bibfnamefont {D.~J.}\ \bibnamefont {Wilkinson}},\ }\href@noop {} {\emph {\bibinfo {title} {Stochastic modelling for systems biology}}}\ (\bibinfo  {publisher} {Chapman and Hall/CRC},\ \bibinfo {year} {2018})\BibitemShut {NoStop}%
\bibitem [{\citenamefont {Mastromatteo}\ \emph {et~al.}(2014)\citenamefont {Mastromatteo}, \citenamefont {Toth},\ and\ \citenamefont {Bouchaud}}]{mastromatteo2014anomalous}%
  \BibitemOpen
  \bibfield  {author} {\bibinfo {author} {\bibfnamefont {I.}~\bibnamefont {Mastromatteo}}, \bibinfo {author} {\bibfnamefont {B.}~\bibnamefont {Toth}},\ and\ \bibinfo {author} {\bibfnamefont {J.-P.}\ \bibnamefont {Bouchaud}},\ }\bibfield  {title} {\bibinfo {title} {Anomalous impact in reaction-diffusion financial models},\ }\href@noop {} {\bibfield  {journal} {\bibinfo  {journal} {Physical Review Letters}\ }\textbf {\bibinfo {volume} {113}},\ \bibinfo {pages} {268701} (\bibinfo {year} {2014})}\BibitemShut {NoStop}%
\bibitem [{\citenamefont {T{\'o}th}\ \emph {et~al.}(2011)\citenamefont {T{\'o}th}, \citenamefont {Lemperiere}, \citenamefont {Deremble}, \citenamefont {De~Lataillade}, \citenamefont {Kockelkoren},\ and\ \citenamefont {Bouchaud}}]{toth2011anomalous}%
  \BibitemOpen
  \bibfield  {author} {\bibinfo {author} {\bibfnamefont {B.}~\bibnamefont {T{\'o}th}}, \bibinfo {author} {\bibfnamefont {Y.}~\bibnamefont {Lemperiere}}, \bibinfo {author} {\bibfnamefont {C.}~\bibnamefont {Deremble}}, \bibinfo {author} {\bibfnamefont {J.}~\bibnamefont {De~Lataillade}}, \bibinfo {author} {\bibfnamefont {J.}~\bibnamefont {Kockelkoren}},\ and\ \bibinfo {author} {\bibfnamefont {J.-P.}\ \bibnamefont {Bouchaud}},\ }\bibfield  {title} {\bibinfo {title} {Anomalous price impact and the critical nature of liquidity in financial markets},\ }\href@noop {} {\bibfield  {journal} {\bibinfo  {journal} {Physical Review X}\ }\textbf {\bibinfo {volume} {1}},\ \bibinfo {pages} {021006} (\bibinfo {year} {2011})}\BibitemShut {NoStop}%
\bibitem [{\citenamefont {Kloeden}\ and\ \citenamefont {Platen}(1992)}]{kloeden1992numerical}%
  \BibitemOpen
  \bibfield  {author} {\bibinfo {author} {\bibfnamefont {P.~E.}\ \bibnamefont {Kloeden}}\ and\ \bibinfo {author} {\bibfnamefont {E.}~\bibnamefont {Platen}},\ }\href@noop {} {\emph {\bibinfo {title} {Numerical Solution of Stochastic Differential Equations}}}\ (\bibinfo  {publisher} {Springer Berlin, Heidelberg},\ \bibinfo {year} {1992})\BibitemShut {NoStop}%
\bibitem [{\citenamefont {Fornberg}(1998)}]{fornberg1998practical}%
  \BibitemOpen
  \bibfield  {author} {\bibinfo {author} {\bibfnamefont {B.}~\bibnamefont {Fornberg}},\ }\href@noop {} {\emph {\bibinfo {title} {A practical guide to pseudospectral methods}}}\ (\bibinfo  {publisher} {Cambridge University Press},\ \bibinfo {year} {1998})\BibitemShut {NoStop}%
\bibitem [{\citenamefont {Frigo}\ and\ \citenamefont {Johnson}(2005)}]{FFTW05}%
  \BibitemOpen
  \bibfield  {author} {\bibinfo {author} {\bibfnamefont {M.}~\bibnamefont {Frigo}}\ and\ \bibinfo {author} {\bibfnamefont {S.~G.}\ \bibnamefont {Johnson}},\ }\bibfield  {title} {\bibinfo {title} {The design and implementation of {FFTW3}},\ }\href@noop {} {\bibfield  {journal} {\bibinfo  {journal} {Proceedings of the IEEE}\ }\textbf {\bibinfo {volume} {93}},\ \bibinfo {pages} {216} (\bibinfo {year} {2005})},\ \bibinfo {note} {special issue on ``Program Generation, Optimization, and Platform Adaptation''}\BibitemShut {NoStop}%
\bibitem [{\citenamefont {Orszag}(1971)}]{orszag1971elimination}%
  \BibitemOpen
  \bibfield  {author} {\bibinfo {author} {\bibfnamefont {S.~A.}\ \bibnamefont {Orszag}},\ }\bibfield  {title} {\bibinfo {title} {On the elimination of aliasing in finite-difference schemes by filtering high-wavenumber components},\ }\href@noop {} {\bibfield  {journal} {\bibinfo  {journal} {Journal of Atmospheric Sciences}\ }\textbf {\bibinfo {volume} {28}},\ \bibinfo {pages} {1074} (\bibinfo {year} {1971})}\BibitemShut {NoStop}%
\bibitem [{\citenamefont {Orszag}(1972)}]{orszag1972comparison}%
  \BibitemOpen
  \bibfield  {author} {\bibinfo {author} {\bibfnamefont {S.~A.}\ \bibnamefont {Orszag}},\ }\bibfield  {title} {\bibinfo {title} {Comparison of pseudospectral and spectral approximation},\ }\href@noop {} {\bibfield  {journal} {\bibinfo  {journal} {Studies in Applied Mathematics}\ }\textbf {\bibinfo {volume} {51}},\ \bibinfo {pages} {253} (\bibinfo {year} {1972})}\BibitemShut {NoStop}%
\bibitem [{\citenamefont {Patterson}\ and\ \citenamefont {Orszag}(1971)}]{patterson1971spectral}%
  \BibitemOpen
  \bibfield  {author} {\bibinfo {author} {\bibfnamefont {G.}~\bibnamefont {Patterson}}\ and\ \bibinfo {author} {\bibfnamefont {S.~A.}\ \bibnamefont {Orszag}},\ }\bibfield  {title} {\bibinfo {title} {Spectral calculations of isotropic turbulence: Efficient removal of aliasing interactions},\ }\href@noop {} {\bibfield  {journal} {\bibinfo  {journal} {Physics of Fluids}\ }\textbf {\bibinfo {volume} {14}},\ \bibinfo {pages} {2538} (\bibinfo {year} {1971})}\BibitemShut {NoStop}%
\bibitem [{\citenamefont {Boyd}(2001)}]{boyd2001chebyshev}%
  \BibitemOpen
  \bibfield  {author} {\bibinfo {author} {\bibfnamefont {J.~P.}\ \bibnamefont {Boyd}},\ }\href@noop {} {\emph {\bibinfo {title} {Chebyshev and Fourier spectral methods}}}\ (\bibinfo  {publisher} {Courier Corporation},\ \bibinfo {year} {2001})\BibitemShut {NoStop}%
\bibitem [{\citenamefont {Mannella}(1997)}]{mannella1997numerical}%
  \BibitemOpen
  \bibfield  {author} {\bibinfo {author} {\bibfnamefont {R.}~\bibnamefont {Mannella}},\ }\bibfield  {title} {\bibinfo {title} {Numerical integration of stochastic differential equations},\ }\href@noop {} {\bibfield  {journal} {\bibinfo  {journal} {Proc. Euroconf. on Supercomputation in Nonlinear and Disordered Systems}\ ,\ \bibinfo {pages} {100}} (\bibinfo {year} {1997})}\BibitemShut {NoStop}%
\bibitem [{\citenamefont {Mannella}(2000)}]{mannella2000gentle}%
  \BibitemOpen
  \bibfield  {author} {\bibinfo {author} {\bibfnamefont {R.}~\bibnamefont {Mannella}},\ }\bibfield  {title} {\bibinfo {title} {A gentle introduction to the integration of stochastic differential equations},\ }in\ \href@noop {} {\emph {\bibinfo {booktitle} {Stochastic Processes in Physics, Chemistry, and Biology}}}\ (\bibinfo  {publisher} {Springer Berlin Heidelberg},\ \bibinfo {year} {2000})\ pp.\ \bibinfo {pages} {353--364}\BibitemShut {NoStop}%
\bibitem [{Note1()}]{Note1}%
  \BibitemOpen
  \bibinfo {note} {See for example the CFD package OpenFOAM, which allows the user to describe each term in a dynamic equation as explicit or implicit: https://www.openfoam.org}\BibitemShut {NoStop}%
\bibitem [{\citenamefont {Zhao}\ \emph {et~al.}(2024)\citenamefont {Zhao}, \citenamefont {Gulati}, \citenamefont {Caballero}, \citenamefont {Kolvin}, \citenamefont {Adkins}, \citenamefont {Marchetti},\ and\ \citenamefont {Dogic}}]{zhao2024asymmetric}%
  \BibitemOpen
  \bibfield  {author} {\bibinfo {author} {\bibfnamefont {L.}~\bibnamefont {Zhao}}, \bibinfo {author} {\bibfnamefont {P.}~\bibnamefont {Gulati}}, \bibinfo {author} {\bibfnamefont {F.}~\bibnamefont {Caballero}}, \bibinfo {author} {\bibfnamefont {I.}~\bibnamefont {Kolvin}}, \bibinfo {author} {\bibfnamefont {R.}~\bibnamefont {Adkins}}, \bibinfo {author} {\bibfnamefont {M.~C.}\ \bibnamefont {Marchetti}},\ and\ \bibinfo {author} {\bibfnamefont {Z.}~\bibnamefont {Dogic}},\ }\bibfield  {title} {\bibinfo {title} {Asymmetric fluctuations and self-folding of active interfaces},\ }\href@noop {} {\bibfield  {journal} {\bibinfo  {journal} {arXiv preprint arXiv:2407.04679}\ } (\bibinfo {year} {2024})}\BibitemShut {NoStop}%
\bibitem [{\citenamefont {Risthaus}\ and\ \citenamefont {Schneider}(2024)}]{risthaus2024imposing}%
  \BibitemOpen
  \bibfield  {author} {\bibinfo {author} {\bibfnamefont {L.}~\bibnamefont {Risthaus}}\ and\ \bibinfo {author} {\bibfnamefont {M.}~\bibnamefont {Schneider}},\ }\bibfield  {title} {\bibinfo {title} {Imposing dirichlet boundary conditions directly for fft-based computational micromechanics},\ }\href@noop {} {\bibfield  {journal} {\bibinfo  {journal} {Computational Mechanics}\ ,\ \bibinfo {pages} {1}} (\bibinfo {year} {2024})}\BibitemShut {NoStop}%
\bibitem [{\citenamefont {Barrata}\ \emph {et~al.}(2023)\citenamefont {Barrata}, \citenamefont {Dean}, \citenamefont {Dokken}, \citenamefont {Habera}, \citenamefont {Hale}, \citenamefont {Richardson}, \citenamefont {Rognes}, \citenamefont {Scroggs}, \citenamefont {Sime},\ and\ \citenamefont {Wells}}]{barrata2023dolfinx}%
  \BibitemOpen
  \bibfield  {author} {\bibinfo {author} {\bibfnamefont {I.~A.}\ \bibnamefont {Barrata}}, \bibinfo {author} {\bibfnamefont {J.~P.}\ \bibnamefont {Dean}}, \bibinfo {author} {\bibfnamefont {J.~S.}\ \bibnamefont {Dokken}}, \bibinfo {author} {\bibfnamefont {M.}~\bibnamefont {Habera}}, \bibinfo {author} {\bibfnamefont {J.}~\bibnamefont {Hale}}, \bibinfo {author} {\bibfnamefont {C.}~\bibnamefont {Richardson}}, \bibinfo {author} {\bibfnamefont {M.~E.}\ \bibnamefont {Rognes}}, \bibinfo {author} {\bibfnamefont {M.~W.}\ \bibnamefont {Scroggs}}, \bibinfo {author} {\bibfnamefont {N.}~\bibnamefont {Sime}},\ and\ \bibinfo {author} {\bibfnamefont {G.~N.}\ \bibnamefont {Wells}},\ }\bibfield  {title} {\bibinfo {title} {Dolfinx: The next generation fenics problem solving environment}\ }\href {https://doi.org/10.5281/zenodo.10447666} {10.5281/zenodo.10447666} (\bibinfo {year} {2023})\BibitemShut {NoStop}%
\bibitem [{\citenamefont {Burns}\ \emph {et~al.}(2020)\citenamefont {Burns}, \citenamefont {Vasil}, \citenamefont {Oishi}, \citenamefont {Lecoanet},\ and\ \citenamefont {Brown}}]{burns2020dedalus}%
  \BibitemOpen
  \bibfield  {author} {\bibinfo {author} {\bibfnamefont {K.~J.}\ \bibnamefont {Burns}}, \bibinfo {author} {\bibfnamefont {G.~M.}\ \bibnamefont {Vasil}}, \bibinfo {author} {\bibfnamefont {J.~S.}\ \bibnamefont {Oishi}}, \bibinfo {author} {\bibfnamefont {D.}~\bibnamefont {Lecoanet}},\ and\ \bibinfo {author} {\bibfnamefont {B.~P.}\ \bibnamefont {Brown}},\ }\bibfield  {title} {\bibinfo {title} {Dedalus: A flexible framework for numerical simulations with spectral methods},\ }\href@noop {} {\bibfield  {journal} {\bibinfo  {journal} {Physical Review Research}\ }\textbf {\bibinfo {volume} {2}},\ \bibinfo {pages} {023068} (\bibinfo {year} {2020})}\BibitemShut {NoStop}%
\bibitem [{Note2()}]{Note2}%
  \BibitemOpen
  \bibinfo {note} {\lowercase {h}ttps://github.com/fcaballerop/cuPSS}\BibitemShut {NoStop}%
\bibitem [{\citenamefont {Caballero}\ \emph {et~al.}(2023)\citenamefont {Caballero}, \citenamefont {You},\ and\ \citenamefont {Marchetti}}]{caballero2023vorticity}%
  \BibitemOpen
  \bibfield  {author} {\bibinfo {author} {\bibfnamefont {F.}~\bibnamefont {Caballero}}, \bibinfo {author} {\bibfnamefont {Z.}~\bibnamefont {You}},\ and\ \bibinfo {author} {\bibfnamefont {M.~C.}\ \bibnamefont {Marchetti}},\ }\bibfield  {title} {\bibinfo {title} {Vorticity phase separation and defect lattices in the isotropic phase of active liquid crystals},\ }\href@noop {} {\bibfield  {journal} {\bibinfo  {journal} {Soft Matter}\ }\textbf {\bibinfo {volume} {19}},\ \bibinfo {pages} {7828} (\bibinfo {year} {2023})}\BibitemShut {NoStop}%
\bibitem [{\citenamefont {Gulati}\ \emph {et~al.}(2024)\citenamefont {Gulati}, \citenamefont {Caballero}, \citenamefont {Kolvin}, \citenamefont {You},\ and\ \citenamefont {Marchetti}}]{gulati2024traveling}%
  \BibitemOpen
  \bibfield  {author} {\bibinfo {author} {\bibfnamefont {P.}~\bibnamefont {Gulati}}, \bibinfo {author} {\bibfnamefont {F.}~\bibnamefont {Caballero}}, \bibinfo {author} {\bibfnamefont {I.}~\bibnamefont {Kolvin}}, \bibinfo {author} {\bibfnamefont {Z.}~\bibnamefont {You}},\ and\ \bibinfo {author} {\bibfnamefont {M.~C.}\ \bibnamefont {Marchetti}},\ }\bibfield  {title} {\bibinfo {title} {Traveling waves at the surface of active liquid crystals},\ }\href@noop {} {\bibfield  {journal} {\bibinfo  {journal} {Soft Matter}\ } (\bibinfo {year} {2024})}\BibitemShut {NoStop}%
\end{thebibliography}%

\end{document}